\documentclass{elsart}

 \usepackage{graphicx}

\usepackage{amssymb}

\newcommand{\nn}{\nonumber}
\newcommand{\smi}{\sum_i}

\newcommand{\yi}{\langle Y_i \rangle}
\newcommand{\yj}{\langle Y_j \rangle}

\newcommand{\xxi}{\langle X_i \rangle}

\newcommand{\smij}{\sum_{j \ne i}}
\newcommand{\smijk}{\sum_{i,j \ne k}}
\newcommand{\cii}{{C}_{ii}}
\newcommand{\cij}{{C}_{ij}}
\newcommand{\vyii}{\langle Y_i^2 \rangle - \langle Y_i \rangle^2}
\newcommand{\vyjj}{\langle Y_j^2 \rangle - \langle Y_j \rangle^2}
\newcommand{\vyij}{\langle Y_i Y_j \rangle - \langle Y_i \rangle
\langle Y_j \rangle}
\newcommand{\vyik}{\langle Y_i Y_k \rangle - \langle Y_i \rangle
\langle Y_k \rangle}
\newcommand{\vyjk}{\langle Y_j Y_k \rangle - \langle Y_j \rangle
\langle Y_k \rangle}
\newcommand{\vxii}{\langle X_i^2 \rangle - \langle X_i \rangle^2}

\newcommand{\bt}{b_{\rm tot}}
\newcommand{\bm}{( M - 1 ) b'}
\newcommand{\eps}{a - b - ( M - 1 ) b'}
\newcommand{\bry}{\left\{ Y_i \right\}}
\newcommand{\bryj}{\left\{ Y_j \right\}}
\newcommand{\vx}{\langle X \rangle}
\newcommand{\vy}{\langle Y \rangle}
\newcommand{\vxx}{\langle X^2 \rangle}
\newcommand{\vyy}{\langle Y^2 \rangle}
\newcommand{\vxy}{\langle X Y \rangle}

\begin{document}
\begin{frontmatter}

\title{Statistics of infections with diversity in the pathogenicity.}
\author[Madrid]{Francisco Guinea}%
\address[Madrid]{Instituto de Ciencia de Materiales de Madrid, CSIC,
Cantoblanco, E-28049 Madrid, Spain.}
\author[London]{Vincent A. A. Jansen}
\author[London]{Nico Stollenwerk}
\address[London]{School of Biological Sciences, Royal Holloway-University
of London, Egham, Surrey TW20 0EX, United Kingdom.}
\begin{abstract}
The statistics of outbreaks in a model for the propagation of
meningococcal diseases is analyzed, taking into account the
possibility that the population is fragmented into weakly
connected patches. It is shown that, depending on the size of the
sample studied, the ration between the variance and the average of
infected cases can vary from unity (Poisson statistics) to
$\epsilon^{-1}$, where $\epsilon$ is the normalized infection
rate.
\end{abstract}
\end{frontmatter}
\maketitle
\section{Introduction.}
The meningococcus is a major cause of meningitis and septicaemia.
Despite this, infection with the meningococcus is mostly harmless
and only rarely leads to disease. Transmission of the disease is
almost exclusively through asymptomatic carriers of the disease. A
predominant feature of the epidemiology of meningococcal disease
are outbreaks of variable scale and duration. The meningococcal
population is genetically highly diverse. We have shown, using a
mathematical model, that heritable diversity with respect to
pathogenic potential can lead to disease outbreaks
\cite{SJ03,SJ03b,SMJ04}.

Meningococcal disease is a a notifiable disease in many countries.
Therefore there exist extensive data sets on the incidence of
meningococcal disease. The analysis of meningococcal disease data
is problematic because the number of asymptomatic carriers at any
time, the variable that is probably of most interest, is normally
not known because transmission of the pathogen takes place almost
exclusively through asymptomatic carriers. Therefore key
epidemiological parameters are difficult to estimate and methods
that are standard in epidemiology, such as outbreak reconstruction
through contact tracing, can not easily be applied. For this
reason outbreaks of meningococcal disease are difficult to
reconstruct and to detect.

In this paper we will investigate the statistical structure of an
epidemiological model to infer the underlying disease process from
data on the number of cases of disease. Such insights have been
applied in the analysis of meningococcal disease data
\cite{SMJ04}. Here we will investigate the validity of the
assumptions made for these inferences and study how the variance
in the number of cases of disease depends on the structure of the
population.

\section{The SIRYX model.}

We study the SIRYX model, considered in \cite{SJ03,SJ03b,SMJ04}.
The model is an extension of the SIR model\cite{AM91}. There are
two types of infected individuals, $I$ and $Y$. The $Y$'s are
generated by mutation from the $I$'s at rate $\mu \beta$. For
simplicity we assume that the back mutation rate $Y \rightarrow I$
is nil. The $Y$ population can develop disease at rate $\epsilon
\beta$. The parameter $\epsilon$ is the pathogenicity: the
probability to develop disease upon infection. We define the
number of individuals which suffer the disease $X$. We further
simplify the model by assuming that these individuals are removed
from the population. The mean field equations are:
\begin{eqnarray}
\frac{d S}{d t} &= &\alpha R - \beta \frac{S}{N_p} ( I + Y )
\nonumber \\ \frac{d I}{d t} &= &\beta(1  - \mu ) \frac{S}{N}I -
\gamma I  \nonumber \\ \frac{d R}{d t} &= &\gamma ( I + Y ) -
\alpha R \nonumber \\ \frac{d Y}{d t} &= &\beta( 1  - \epsilon )
\frac{S}{N_p} Y - \gamma Y + \beta \mu \frac{S}{N_p} I \nonumber
\\ \frac{d X}{d t} &= &\beta \epsilon \frac{S}{N_p} Y - \beta \delta X\label{SIRYX}
\end{eqnarray}

The only difference with respect to the model studied
in\cite{SJ03,SMJ04} is the introduction of a the rate $\delta$ at
which the $X$ individuals are removed from the population (see
below). This rate implies, that, in the long run, the only
stationary situation is the conversion of all individuals into the
$X$ type, and their eventual disappearance. We will study here
quasisatationary situations which arise when $\delta , \epsilon
\ll 1$.

Following \cite{SJ03}, we consider that the system is near its
stable point when $\mu=0$ and $Y=0$. The remaining parameters at
the fixed point are:
\begin{eqnarray}
\frac{S}{N_p} &= &\frac{S^*}{N_p} = \frac{\gamma}{\beta} \nonumber
\\ \frac{I}{N_p} &= &\frac{I^*}{N_p} = \frac{\alpha}{\beta}\frac{\beta
- \gamma}{\alpha + \gamma} \label{critical} \end{eqnarray}

Assuming that the fixed point values for $S$ and $I$ do not change
much for small $\mu$, we can define a simple birth-death model for
the variables $Y$ and $X$. We define $p ( Y , t ) =
\sum_{X=0}^{\infty} p (Y , X , t )$, as the probability of finding
the value $Y$ at time $t$. This function satisfies:
\begin{eqnarray}
\frac{d}{d t} p ( Y , t ) &= &[ b ( Y - 1 ) + c ] p ( Y-1 , t )
\nn \\ &+ &a ( Y+ 1 ) p ( Y + 1 , t ) - ( b Y + a Y + c ) p ( Y ,
t ) \label{partial}
\end{eqnarray}
where we have defined the death rate, $a = \gamma$, birth rate, $b
= \beta (1- \epsilon) S^* / N_p$ and rate of creation of a new
individuals by  mutation, $c = \beta \mu ( S^* / N_p ) I^*$.

We generalize this equation to the case of a system divided into
$M$ patches. The main difference is that the birth probability has
to be divided into the probability that the contagion is to
another individual within the same patch, which we still define as
$b$, and the probability that the contagion leads to a new
individual of type $Y$ in another patch, $b'$. The total infection
rate remains $ b + \bm=\beta (1- \epsilon) S^* / N_p$. The
generalization of eq. (\ref{partial}) is:
\begin{eqnarray}
\frac{d}{d t} p ( \bry , t ) &= &\smi b ( Y_i - 1 ) p ( Y_i - 1 ,
\bryj , t ) +  \smij \smi b' Y_j p ( Y_i - 1 , \bry , t ) +\nn \\
&+ &\smi a ( Y_i + 1 ) p ( Y_i + 1 , \bryj , t )  + \smi c p ( Y_i
- 1 , \bryj , t ) - \nn \\ &- &\smi b Y_i p ( \bry , t ) - \smij
\smi b' Y_j p ( \bry , t ) - \nn \\ &- & \smi a Y_i p ( \bry , t )
- c p ( \bry , t ) \label{partial_2}
\end{eqnarray}
Note that in this equation $c$ is the mutation rate within one
patch. The total mutation rate is $M c$. When $b = b'$ we recover
the limit of a well mixed population, while for $b' = 0$ the
patches are decoupled.
\section{Results.} From eq.(\ref{partial_2}) we can calculate the ensemble
means of different quantities. The details of the calculations are
given in the Appendix. The results are:
\begin{eqnarray}
\yi &= &\frac{c}{\eps} \nonumber \\
\vyii &= &\frac{ac [ a - b - ( M - 2 ) b' ]}{[ \eps ]^2 ( a - b +
b' )}
\nonumber \\
 \vyij &= &\frac{a b' c}{[ \eps ]^2 ( a - b + b'
)} \label{averages}
\end{eqnarray}
All these quantities vanish when the mutation rate is zero, $c=0$.

The net growth rate is $a \epsilon = \eps$. The number of infected
cases appear with rate $X'_i=a \epsilon Y_i$.

We now calculate the number of infected individuals, $X_i$. We
study first the case of a single population and a single variable
$X$. The infected individuals are generated from the $Y$'s at rate
$a \eps$. In order to calculate $X$ in a single population, we use
as unit of time $a^{-1}$, and assume that the death rate of the
$X$'s is $\delta$. We write the mutation rate $c = a \gamma$.
then, we can write:
\begin{eqnarray}
\frac{d P ( X , Y , t )}{d t} &= &\epsilon \left[ ( Y + 1 ) P ( Y
+ 1 , X - 1 , t ) - Y P ( X , Y , t ) \right] + \nn \\ &+ &\delta
\left[ ( X + 1 ) P ( X + 1 , Y , t ) - X P ( X , Y , t ) \right] -
\gamma P ( X , Y , t ) \label{ev_X} \end{eqnarray} where we have
used as the unit of time $a^{-1}$, $\epsilon$ is now the rate of
conversion from $Y$ into $X$, $\delta$ is the death rate of the
$X$'s, and $\gamma = c / a$ is the mutation rate from $I$ into
$Y$. Using the techniques described in the Appendix, we can write:
\begin{eqnarray}
\frac{d \vx}{d t} &= &\epsilon \vy - \delta \vx \nonumber \\
\frac{d \vxx}{d t} &= &2 \epsilon \vxy + \epsilon \vy - 2 \delta
\vxx + \delta \vx \nonumber \\
\frac{d \vxy}{d t} &= &- \epsilon \vxy + \epsilon \vyy - \epsilon
\vy - \delta \vxy + c \vx
\end{eqnarray}
In a stationary state the right hand side of these equations is
equal to zero, and we find:
\begin{eqnarray}
\vx &= &\frac{\epsilon}{\delta} \vy \nonumber \\
\vxx &= &\frac{\vx}{2} + \frac{\epsilon}{\delta} \left( \vxy +
\frac{\vy}{2} \right) \nonumber \\ \vxy &= &\frac{\epsilon \vyy -
\epsilon \vy + c \vx}{\epsilon + \delta} \end{eqnarray} We
substitute the first and third of these equations into the second,
so that:
\begin{eqnarray}
\vxx - \vx^2 &= &\frac{\epsilon}{\epsilon + \delta} \vy +
\frac{\epsilon^2}{\delta ( \epsilon + \delta )} \left( \vyy -
\vy^2 \right) - \nn \\ &- &\frac{\epsilon^3}{\delta^2 ( \epsilon +
\delta )} \vy^2 + \frac{\vx}{2} + \frac{\epsilon \gamma}{\delta (
\epsilon + \gamma )} \vx \label{av_X} \end{eqnarray} From
eq.(\ref{ev_X}) we also obtain $\vy = \gamma / \epsilon$.
Inserting  this result into eq.(\ref{av_X}), we have:
\begin{equation}
\frac{\vxx - \vx^2}{\vx} = \frac{1}{2} + \frac{\delta}{\epsilon +
\delta} + \frac{\epsilon}{\epsilon + \delta} \frac{\vyy -
\vy^2}{\vy} \label{var_X} \end{equation} This equation relates the
variance and the average of $X$. When the mortality rate is very
high, $\delta \gg \epsilon$, we have:
\begin{equation}
\left. \frac{\vxii}{\xxi} \right|_{ \delta \gg \epsilon} \approx 1
\end{equation}
The ratio approaches a constant of order unity, and the process
seems to have Poisson statistics. This is reasonable, because
there is an approximately constant reservoir of $Y$ individuals
which can lead to an $X$ individual which disappears quickly, and
the distribution of $X$ cases is not influenced by the
fluctuations of $Y$.

A more interesting regime arises if $\delta \ll \epsilon$ and
$\epsilon \ll 1$. Then, the r.h.s. in eq.(\ref{var_X}) is
dominated by the third term, because $( \vyy - \vy^2 ) / \vy \sim
\epsilon^{-1}$. We find in this case:
\begin{equation}
\left. \frac{\vxii}{\xxi} \right|_{\delta \ll \epsilon \ll 1}
\approx \frac{\vyii}{\yi} \label{var_final}
\end{equation}
This result is the basis of the following section. Note that when
$\delta = 0$ the value of $\xxi$ increases linearly with time.
\section{Size effects.}
Using the results in the Appendix and eq.(\ref{var_final}) we find
(for $M \ge 2$):
\begin{equation} \frac{{\rm Var} X_i}{\langle X_i \rangle} =
\frac{a [ a - b - ( M - 2 ) b' ]}{[ \eps ] ( a - b + b' )}
\label{varxi}
\end{equation} On the other hand, for the entire system we obtain:
\begin{equation} \frac{{\rm Var}
X}{\langle X \rangle} = \frac{a }{\eps} \label{varx}
\end{equation}
The linear relationship between the variance and the mean is
discussed in detail in\cite{SMJ04}.
\begin{figure}[!]
\resizebox{10cm}{!}{\includegraphics[]{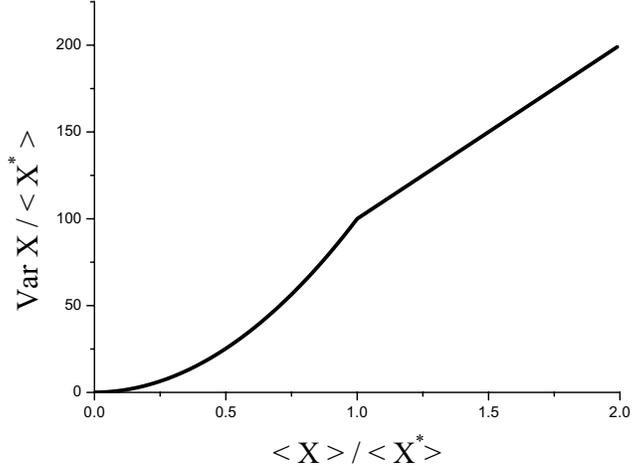}}
\caption{Dependence of the variance of infected individuals on the
average for small populations which are part of a larger, well
mixed population, $\langle X \rangle < \langle X^* \rangle$, or
form an isolated population, $\langle X \rangle \ge \langle X^*
\rangle$ (see eq.(\protect{\ref{varN}})). The infection rate is
$\epsilon = 0.01$.} \label{stat}
\end{figure}
For isolated patches, $b' = 0$ and $b/a = 1 - \epsilon$. As
expected, the local and global values, eq.(\ref{varxi}) and
eq.(\ref{varx}) coincide, giving a ratio equal to $1 / \epsilon$.

In a well mixed population, we have $b' = b$, the total birth rate
is $\bt = M b$, and $( M b ) / a = 1 - \epsilon$. Then, we find:
\begin{eqnarray}
\frac{{\rm Var} X_i}{\langle X_i \rangle} &= &1 + \frac{1 -
\epsilon}{M \epsilon} \label{var_par} \\ \frac{{\rm Var}
X}{\langle X \rangle} &= &\frac{1}{\epsilon} \label{var_tot}
\end{eqnarray} For a small subsystem of a well mixed population
($M \gg 1 / \epsilon$), we have ${\rm Var} X_i / \xxi \approx 1$.
This ratio would imply that the process is due to random mutations
with Poisson statistics. An analysis of the total variance,
however, gives a rather different result. For large (but
artificial) subdivisions of the well mixed system, $M \epsilon \ll
1$, and ${\rm Var} X_i / \xxi \approx 1 / ( M \epsilon )$.

It is interesting to analyze the situation in which populations of
size $N$  below some size $N^*$ are part of a well mixed
population of size $N^*$, while larger populations can be
considered as isolated. made up of smaller, decoupled populations
of size $N^*$. Then, for populations $N \le N^*$ we can use
eq.(\ref{var_par}) with $M = N^* / N = \langle X \rangle / \langle
X^* \rangle$ ($\langle X^* \rangle$ is the value of the mean of a
population of size $N^*$), while when $N \ge N^*$ we can use
eq.(\ref{var_tot}). The variance can be written as:
\begin{equation} {\rm Var} X = \left\{
\begin{array}{lr} \langle X \rangle + \frac{\langle X \rangle^2 ( 1 -
\epsilon )}{\langle X^* \rangle \epsilon} &N < N^* \\
\frac{\langle X \rangle}{\epsilon} &N \geq N^* \end{array} \right.
\label{varN}
\end{equation} . Eq.(\ref{varN}) interpolates
between a Poisson like regime for $N \ll N^*$ to a $1/\epsilon$
ratio between the variance and the mean for $N \geq N^*$. A sketch
of the results is shown in Fig.[\ref{stat}]. 
used here imply that the coupling between different parts of the
\section{Acknowledgements.}
One of us (FG) is thankful to the Royal Society for a travel
grant, and to the Royal Holloway for hospitality. F. G. also
acknowledges financial support from grant MAT2002-0495-C02-01,
MCyT (Spain). V. A. A. J. and N. S. acknowledge financial support
from The Wellcome Trust Grant 063143.
\appendix
From eq.(\ref{partial_2}) one finds the
equations:

\begin{eqnarray}
\frac{ d \yi}{d t} &= &( b - a ) \yi + b' \smij \yj + c \nn \\
\frac{d \vyii}{d t} &= &2 ( b - a ) \left( \vyii \right) + ( a + b
) \yi + c + 2 b' \smij \yj + b' \smi \yi \nn \\ \frac{d \vyij}{d
t} &= &2 ( b - a ) \left( \vyij \right) + \nn \\ &+ &b' \smijk
\vyik + \vyjk + \nn
\\
&+ &b' \smij \vyii + \vyjj \label{averages_2}
\end{eqnarray}
so that $\lim_{t \rightarrow \infty}
\yi = K e^{- [ a - b - \bm ] t} + c / [ \eps ]
$ where $K$ is a constant determined by the initial conditions. We
define:

\begin{eqnarray}
\cii &= &\vyii \nn \\ \cij &= &\vyij \end{eqnarray} These
quantities satisfy: 
\begin{eqnarray} \frac{d}{d t} \left( \begin{array}{c}
\cii \\ \cij \end{array} \right) &= &\left( \begin{array}{cc} 2 (
b - a ) & 2 \bm \\ 2 b' & 2 ( b - a ) + ( M - 2 ) b'
\end{array} \right) \left(
\begin{array}{c} \cii \\ \cij \end{array} \right) + \nn \\ &+ &\left(
\begin{array}{c} ( a + b ) \yi + b' \smij \yj + c \\ 0 \end{array}
\right) \label{matrix}\end{eqnarray} 
At long times, we find:
\begin{equation}
\lim_{t \rightarrow \infty} \left[ ( a + b ) \yi ( t ) + b' \smij
\yj ( t ) + c \right] = \frac{2 a c}{\eps} \end{equation} This
leads to:
\begin{equation}
\lim_{t \rightarrow \infty} \left( \begin{array}{c} \cii \\ \cij
\end{array} \right) = - \left( \begin{array}{cc} 2 ( b
- a ) & 2 \bm \\ 2 b' & 2 ( b - a ) + ( M - 2 ) b'
\end{array} \right)^{-1} \left(
\begin{array}{c} \frac{2 a c}{\eps} \\ 0 \end{array} \right)
\end{equation} 
\end{document}